# X-ray-induced electrical conduction in the insulating phase of thiospinel $CuIr_2S_4$

Takao Furubayashi [a] *, Hiroyuki Suzuki [a], Takehiko Matsumoto [a], and Shoichi Nagata [b]

[a] National Institute for Materials Science, 1-2-1 Sengen, Tsukuba 305-0047, Japan

[b] Department of Materials Science and Engineering, Muroran Institute of Technology, 27-1 Mizumoto-cho, Muroran 050-8585, Japan

Abstract

Effects of x-ray irradiation on the crystal structure and the electrical resistance were examined at low temperatures for the insulating phase of spinel compound $CuIr_2S_4$. We found that the resistance decreases by more than five decades by irradiation at 8.5 K. The structural change from triclinic to tetragonal was observed at the same time. The x-ray-induced conductance is deduced to result from the destruction of $Ir^{4+}$ dimers formed in the insulating phase. Slow relaxation of the resistance in the x-ray-induced state is also reported.

* Corresponding author.

Tel: +81 29 859 2754, Fax: +81 29 859 2701

E-mail: furubayashi.takao@nims.go.jp

Chalcogenide spinel compound $CuIr_2S_4$ has the normal spinel structure with Cu in the *A* site, which is tetrahedrally coordinated with S, and Ir in the *B* site, which is octahedrally coordinated. The compound is metallic around room temperature and exhibits a first-order transition to an insulating phase with the triclinic structure at the temperature $T_{MI} \cong 230$ K. [1-5] The resistivity increases by more than three decades at the transition and becomes semiconducting below $T_{MI}$. Measurements of magnetic susceptibility show that the compound is Pauli paramagnetic in the metallic phase. The susceptibility decreases at the transition and becomes diamagnetic, indicating the disappearance of the Fermi surface at the transition. Results of Cu NMR, [6] XPS [7] and band calculations [8] suggest that Cu atom has almost filled 3d orbital resulting in the $Cu^{1+}$ state. It seems most probable that the ionic configuration of $Cu^{1+}Ir^{3+}Ir^{4+}S^{2-}_4$ and the ordering of two anions, $Ir^{3+}$ and $Ir^{4+}$, are realized as the insulating state [2, 7]. In addition, magnetization measurements and NMR indicates that the insulating phase is diamagnetic. The $Ir^{4+}$ ion is expected to have a spin of $S$=1/2 with the electronic configuration of $(5d\varepsilon)^5$ while $Ir^{3+}$ is in the state $S$=0 with $(5d\varepsilon)^6$. Therefore, the $Ir^{4+}$ possibly forms dimers resulting in the spin singlet state.

In the first report of the structural analysis, a tetragonal structure with the space group $I4_1/amd$ was proposed for the insulating phase [2]. The structure is obtained by expanding the cubic unit cell of spinel along one of the unit vector and shortening along the other two. Although the structure can explain strong reflections, there were some unresolved superstructure reflections. Recently, Radaelli *et al.* [5] have proposed a triclinic structure with the space group $P\bar{1}$ for explaining the superstructure reflections. In this model, the Ir sites are assigned to the same number of $Ir^{3+}$ and $Ir^{4+}$. The $Ir^{4+}$ ions forms pairs with the distance shorter than other Ir-Ir nearest-neighbor distance by more than 10 %. Thus, the structure is consistent with the charge separation model with the formation of $Ir^{4+}$ dimers.

In addition, a further structural change, extinction of the superstructure spots, has been observed by electron diffraction below 50K [3]. Ishibashi *et al.* [9] showed that the transition to tetragonal with the distinction of the superstructure spots is induced by irradiation with x-ray at low temperatures. In addition, electrical resistance was found to decrease at the same time.

This paper reports extensive studies of the effects of x-ray irradiation on the electrical and structural properties of the insulating phase of $CuIr_2S_4$. The decrease of the electrical resistance by more than five decades was observed by illuminating with x-ray. Very slow relaxation in the resistance was found after irradiation, showing the metastable characteristics of the x-ray-induced state. The mechanism of the irradiation-induced conduction is discussed in the relation with the dimer model of the metal-insulator transition.

Samples were prepared in the solid reaction method described previously. [1,2] Powder x-ray diffraction was measured by using a conventional diffractometer equipped with a curved-graphite monochromator on the counter side. A Cu K$\alpha$ radiation was used for the measurements. Low temperatures down to 8.5 K were attained by using a closed-cycle helium refrigerator. The sample powder was pressed onto a copper plate coated slightly with Apiezon N grease and the copper plate was attached to the cold part of the refrigerator. Diffraction patterns were obtained by conventional $\theta$-$2\theta$ scans. Electrical resistance under x-ray irradiation was measured with the conventional DC 4-terminal or 2-terminal method by using the same cryostat and the x-ray apparatus for the diffraction measurements. Electric leads were attached with silver paste on the surface of a sintered polycrystalline sample. An x-ray from the Cu target with the excitation of 50 kV and 300mA was used for irradiating the sample. The direction of the x-ray was perpendicular to the surface where the electrodes were attached. No monochromator was used for the irradiating x-ray. The flux on the sample is estimated to be $3\times10^{12}$ photons/cm$^2\cdot$sec.

The diffraction pattern at room temperature was well reproduced by the normal spinel structure with the space group $Fd\bar{3}m$ and the lattice constants, $a$ = 9.8536 Å and $u$ = 0.387. Figure 1 (a) shows the x-ray diffraction patterns at 160 K. The diffraction profile below $T_{MI}$ was well refined by the triclinic structure [5]. In this paper, the indices for the insulating phase are shown with respect to the cubic spinel unit cell for simplicity, although this is different from the true unit cell of the triclinic structure. The structure is approximately expressed by the tetragonal structure obtained by expanding the cubic lattice along [001] direction and shortening along [100] and [010] of the cubic unit cell. In addition, the superstructure spots are observed as shown in the figure. Then the sample was cooled to 8.5 K and exposed to x-ray for 6 hours. Then the extra spots became almost undetectable as shown in Fig. 1(b). The pattern was well reproduced by the tetragonal structure with the space group $I4_1/amd$.

The change of the diffraction pattern is shown in Fig. 2 for the temperature and the irradiation time. After the pattern (a) was recorded at 160 K, the sample was cooled to 8.5 K and kept for 16 hours at this temperature. The pattern (b) taken at 8.5 K after that shows no significant change with that at 160 K. However, we found that the intensity of the superstructure lines, (1 2 2) and (5/2 1/2 3/2), becomes rapidly weak by increasing the exposing time of the x-ray at 8.5 K as shown in Fig. 2(c)-(f). With increasing the temperature, the superstructure reflections appear at about 80 K and fully recover at 120 K as shown in Fig. 2(g)-(j).

The resistivity of the sintered sample was $4.2\times10^{-3}$ Ω·cm at room temperature. The resistance measured while cooling is shown in Fig. 3 against the temperature $T$. As reported previously [1,2], the resistance increased by more than three decades at $T_{MI}$. The resistance below $T_{MI}$ is well expressed by the form proportional to $\exp(T/T_0)^{-\alpha}$ [10], derived from the mechanism of variable range hopping [11]. Depending on the temperature range, the data can be fitted with $\alpha$=1/4, expected for hopping by phonons [11], or $\alpha$=1/2 for the case of strong

Coulomb interaction [12]. A plot against $T^{-1/2}$ is shown in Fig. 4. The data showed no good fit to the conventional activation-type form, $\exp(T/T_0)^{-1}$. At the lowest temperature of 8.5 K, the resistance is too high to be measured by our system and it is estimated to be larger than $10^{11}$ Ω. Then, the resistance was measured while the sample was irradiated with x-ray at 8.5 K. Figure 5 shows the resistance against the time after starting the irradiation. We found that the resistance decreased drastically to the order of $10^5$ Ω, by more than five decades, after irradiating for about 100 minutes. The sample persists to be conductive after stopping the x-ray, although the resistance shows some slow increase as shown in the inset of Fig. 5.

Temperature dependence of the resistance after the irradiation is also shown in Fig. 3. The data were recorded while warming after waiting for 16 hours at 8.5 K. With increasing the temperature, the resistance once decreases and increases above 30 K. The value recovers to the curve taken on cooling at about 80K. The temperature is lower than 120 K for the full recovery of the superstructure reflections in the diffraction patterns shown in Fig. 2. This is presumably because the structure is still strongly disordered in this temperature range and the superstructure reflections become broad.

The resistance increased after stopping the irradiation at 8.5 K as shown in Fig. 6. The sharp increase in a short time is presumably because of the temperature change by illuminating with the x-ray. Subsequently, we observed a very slow increase of the resistance. For showing the relaxation behavior, the data was re-plotted in Fig. 6 against $\log(t-t_0)$, where $t$ is the time and $t_0$ is defined as the time 10 minutes after stopping the irradiation. We also show the curves at other temperatures in this figure. The sample was cooled to 8.5 K from 160 K and irradiated with the x-ray for 6 hours. Then it was warmed to each indicated temperature. For each temperature, $t_0$ was defined as the time after waiting for 10 minutes to stabilize the temperature. The relaxation curves did not fit to a simple exponential form. Apparently, the relaxation becomes faster with increasing the temperature. However, no sign of saturation is observed at each temperature in the range of the measurement time.

It is to be noticed that the conductive phase would be formed with the thickness comparable with the penetration depth of the x-ray. The penetration depth for the most intense Cu-K$\alpha$ radiation with the energy of 8.05 keV is calculated to be 8.2 μm. The resistivity of the conductive phase at 8.5 K is calculated to be 65 Ω·cm by assuming the thickness equal to the penetration depth. The value might be somewhat underestimated because the penetration depth would be larger due to white x-ray with the energy up to 50 keV. The decrease of the resistivity by irradiation is more than seven decades. The resistivity is still much larger than that of the metallic phase at room temperature, $4.2\times10^{-3}$ Ω·cm. In addition, the result shows clearly the negative temperature coefficient for the resistance after the irradiation. Thus, the tetragonal phase cannot be still called metallic, but it is semiconducting.

The obtained resistivity of the x-ray-induced phase, 65 Ω·cm, is somewhat larger than the value 2 Ω·cm reported by Ishibashi *et al.* [9]. However, the decrease of the resistance by irradiation, more than five decades, is much larger in this work. Ishibashi *et al.* reported that the resistance decreased by less than a decade (three decades in the resistivity). The difference comes from the result that the resistivity of the triclinic phase at the lowest temperatures is much larger in this work. Apparently, the difference could be attributed to the quality of the sample. Although it is not readily concluded which is better, it is to be mentioned that the resistivity of the insulating phase at low temperatures decreases by introducing air while the preparation process [10].

Ishibashi *et al.* [9] concluded that the observed tetragonal phase intrinsically has the triclinic lattice distortion in the short-range scale from the analysis of the diffuse reflections. They attributed the electrical conductance in the irradiated state to charge disorder at the boundaries of the triclinic grains. It is unclear, however, how such disordered state leads to the electric conduction. The localized electrons of the dimers would not contribute the conduction, even if the spatial configuration of the dimers is strongly disordered.

We propose the following explanation for the x-ray-induced conduction. In the triclinic insulating phase, $Ir^{4+}$ pairs form dimers with the distance remarkably smaller than any other combinations. The characteristic superstructure reflections appear as the result of the lattice distortion due to the dimer formation. It is reasonable to consider that an attractive force, which originates from the formation of the spin-singlet state, is working between the $Ir^{4+}$ ions. As pointed by Ishibashi *et al.*, the x-ray irradiation can cause the disorder of the charge by the photoelectron effect, similarly with the x-ray-induced transition in magnetoresistive manganite [13, 14]. One possibility is that an electron is removed from 5dγ orbital and $Ir^{4+}$ turns to $Ir^{5+}$. It would be also possible that an electron from an excited state fills the 5dγ orbital of $Ir^{4+}$. Although the detailed process is unclear, the attractive force between the $Ir^{4+}$ ions disappears in any case and the Ir atoms would move to the equally spaced position. The ionic state would recover to $Ir^{4+}$ in a short time because of the charge neutrality. However, the $Ir^{4+}$ pair cannot readily form a dimer at low temperatures because a potential barrier should be surmounted in moving to the attracted position. The presence of such a potential barrier is reasonable in considering that the MI transition is of the first order.

Thus, the x-ray irradiation can produce $Ir^{4+}$ that do not form dimers. According to Ishibashi *et al.* [9], a considerable number of $Ir^{4+}$ ions still form dimers in the x-ray-induced phase. However, the presence of non-dimerized $Ir^{4+}$ is not excluded. Because an $Ir^{4+}$ ion has one less electron in the 5dγ orbital with respect to $Ir^{3+}$, it can be considered that $Ir^{4+}$ has a hole. The hole of $Ir^{4+}$ is not locally bounded but it can move to neighboring $Ir^{3+}$ ions, leading to conductive state. The presence of the potential barrier is consistent with the slow relaxation observed in the electrical resistance. This implies that the conductive state relaxes to the insulating phase with a very long time constant. As shown in Fig. 5, the relaxation is faster with increasing the temperature, indicating thermally assisted process to surmount the potential barrier to form a dimer. The relaxation would become fast enough compared with the measuring time at about 80 K, where the thermal excitation becomes comparable with the barrier height. Thus, the resistivity is recovered to the value before the irradiation.

Metal-insulator transitions in doped semiconductors have been observed in a variety of systems [15]. If carriers more than a certain critical concentration are doped in a semiconductor, the system shows a metallic conduction. In this work, we have discussed that the destruction of $Ir^{4+}$ dimers by x-ray leads to doping holes. The negative temperature coefficient of the resistance after irradiation indicates the semiconducting characteristics of the x-ray-induced state. Thus, the holes doped by destructing $Ir^{4+}$ dimers would not be still enough to cause a transition to a metallic state.

Finally, we discuss the mechanism of the electrical conduction of the triclinic insulating phase before irradiation. As shown in Fig. 4, the temperature dependence of the resistance at low temperatures is typical of variable range hopping. In general, activation-type temperature dependence is observed in a crystalline semiconductor with a well-defined energy gap. Thus, variable range hopping is the evidence that there is some randomness in the conduction mechanism. In the light of the dimer model, we can deduce that the possible randomness in the formation of $Ir^{4+}$ dimers would be the origin of the conduction. Lattice defects such as atomic vacancies can disturb the formation of dimers and produce non-dimerized $Ir^{4+}$ ions. Such non-dimerized ions randomly distributed in the sample cause variable range hopping through doping holes. Thus, on the basis of the dimer model, a quite similar argument with the x-ray-induced conduction is applicable to the conduction of the insulating phase.

To summarize, effects of x-ray irradiation on the electrical resistance was examined for the insulating phase of $CuIr_2S_4$ at low temperatures. We observed the decrease of the resistance by more than five decades by irradiation at 8.5 K. At the same time, the crystal structure was transformed to tetragonal with the extinction of the superstructure lines, which are evidenced for the $Ir^{4+}$ dimers in the insulating phase with the triclinic structure. It can be deduced that the $Ir^{4+}$ dimers are destroyed by the photoelectron effect and supply holes, which cause the enhancement of the electrical conductance. The $Ir^{4+}$ dimers, after once destroyed, are not readily formed again at low temperatures because of a potential barrier to be surmounted to cause the lattice deformation necessary in forming dimers. The slow relaxation observed in the electrical resistance is consistent with the presence of the potential barrier.

Acknowledgements

The authors would like to thank T. Sasaki and A. Fujimori for fruitful discussions.

London, 1990, p.145.

Figure Captions

Fig. 1. Powder x-ray diffraction at each temperature: (a) after cooled to 160 K from room temperature (multiplied by 10), (b) after irradiated by x-ray for 6 hours at 8.5 K. The indices are for the cubic unit cell. The indices marked by asterisks are superstructure reflection.

Fig. 2. X-ray diffraction patterns in the following conditions. Recorded at (a) 160 K and (b) at 8.5 K after kept for 16 hours at this temperature. Then, patterns were recorded after irradiation with x-ray in the direction perpendicular to the surface at 8.5 K for (c) 30min, (d) 60min, (e) 90 min and (f) 360 min. After stopping the x-ray, the sample was warmed to (g) 40 K, (h) 80 K, (i) 120 K and (j) 160 K.

Fig. 3. Temperature dependence of the electrical resistance. Open triangles were recorded with decreasing temperature from room temperature. Close circles indicate the values taken with increasing the temperature after x-ray irradiation at 8.5 K.

Fig. 4. The resistance before (open triangles) and after (close circles) irradiation, the same data with Fig. 3, is plotted against $T^{-1/2}$, where $T$ is the temperature.

Fig. 5. The resistance at 8.5 K is shown against the irradiation time. The x-ray was stopped at the time indicated by the arrow. The inset shows the enlarged plot for the time larger than $3\times10^4$ sec.

Fig. 6. Time dependence of the normalized resistance of the irradiation-induced state recorded at each temperature. See the text for the measurement conditions.

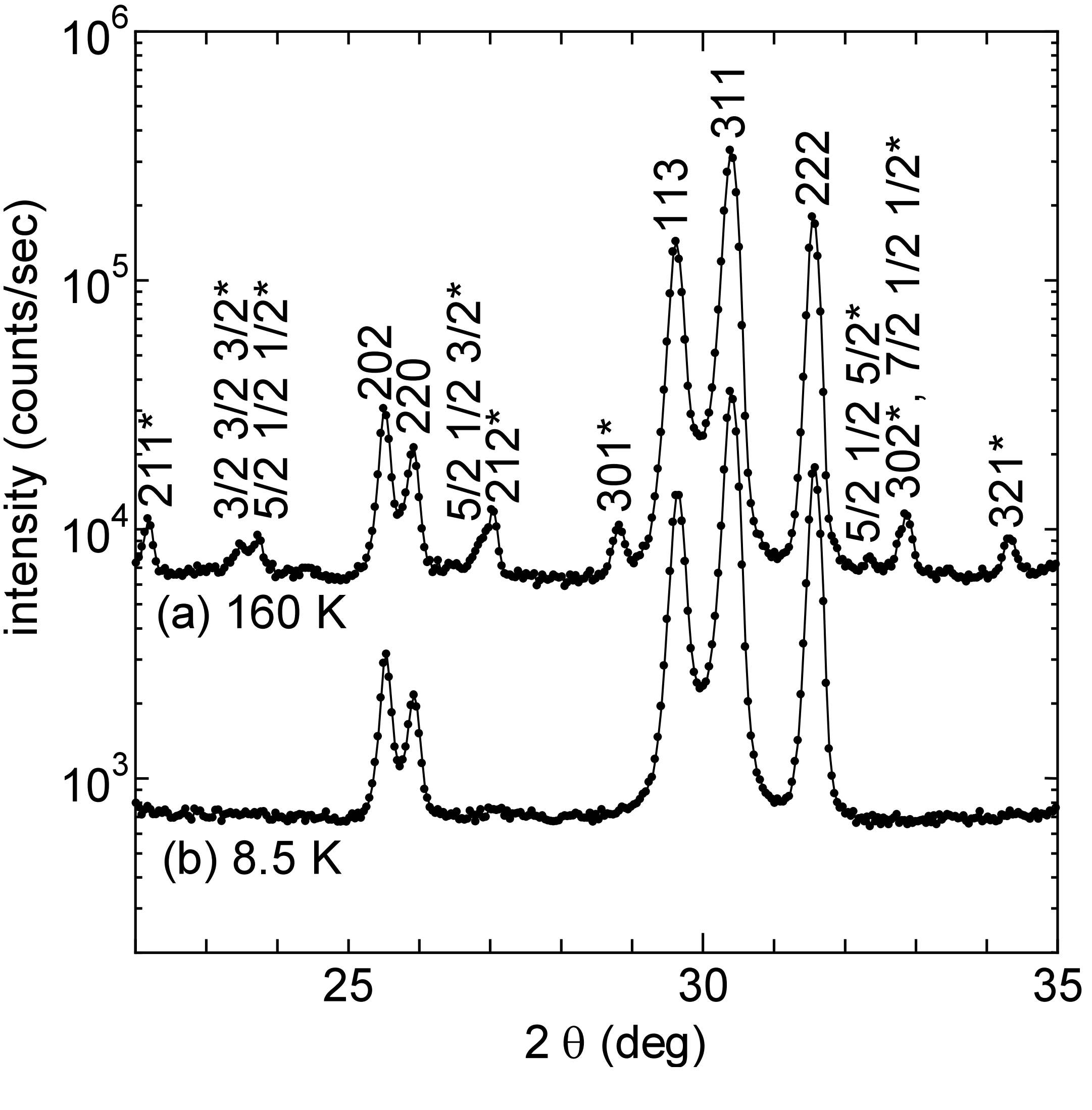

Fig.1, T. Furubayashi

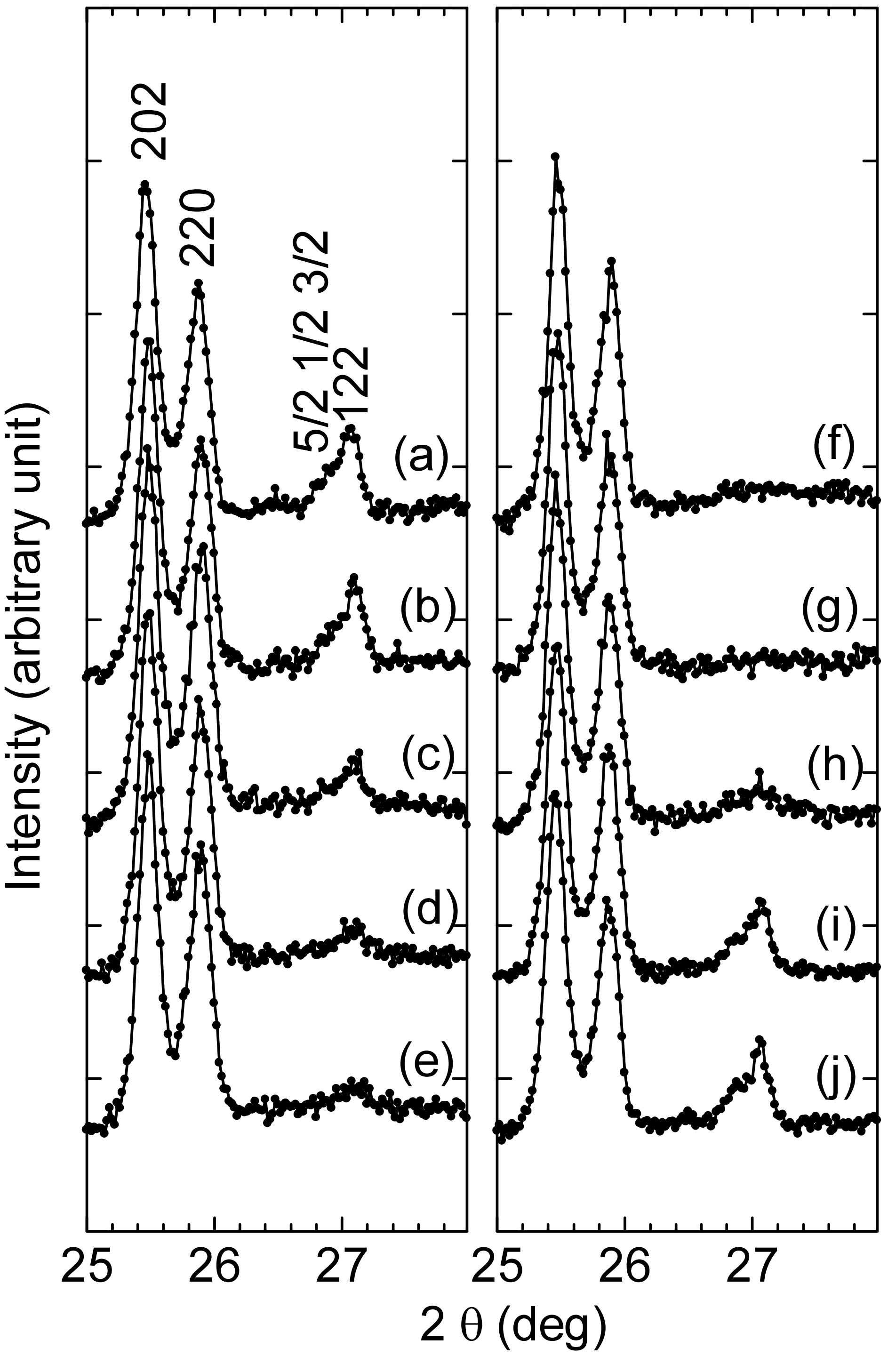

Fig.2, T. Furubayashi

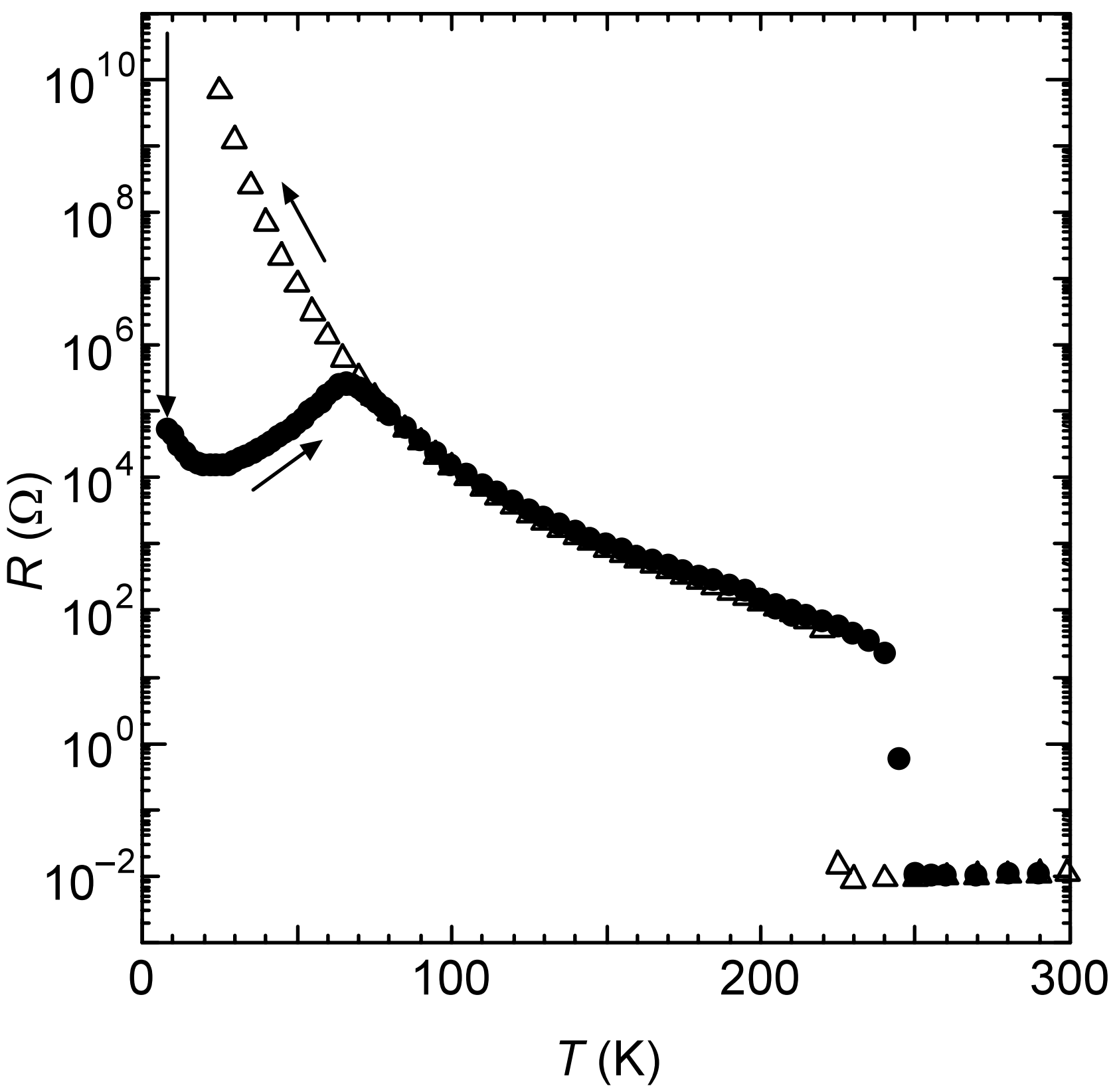

Fig.3, T. Furubayashi

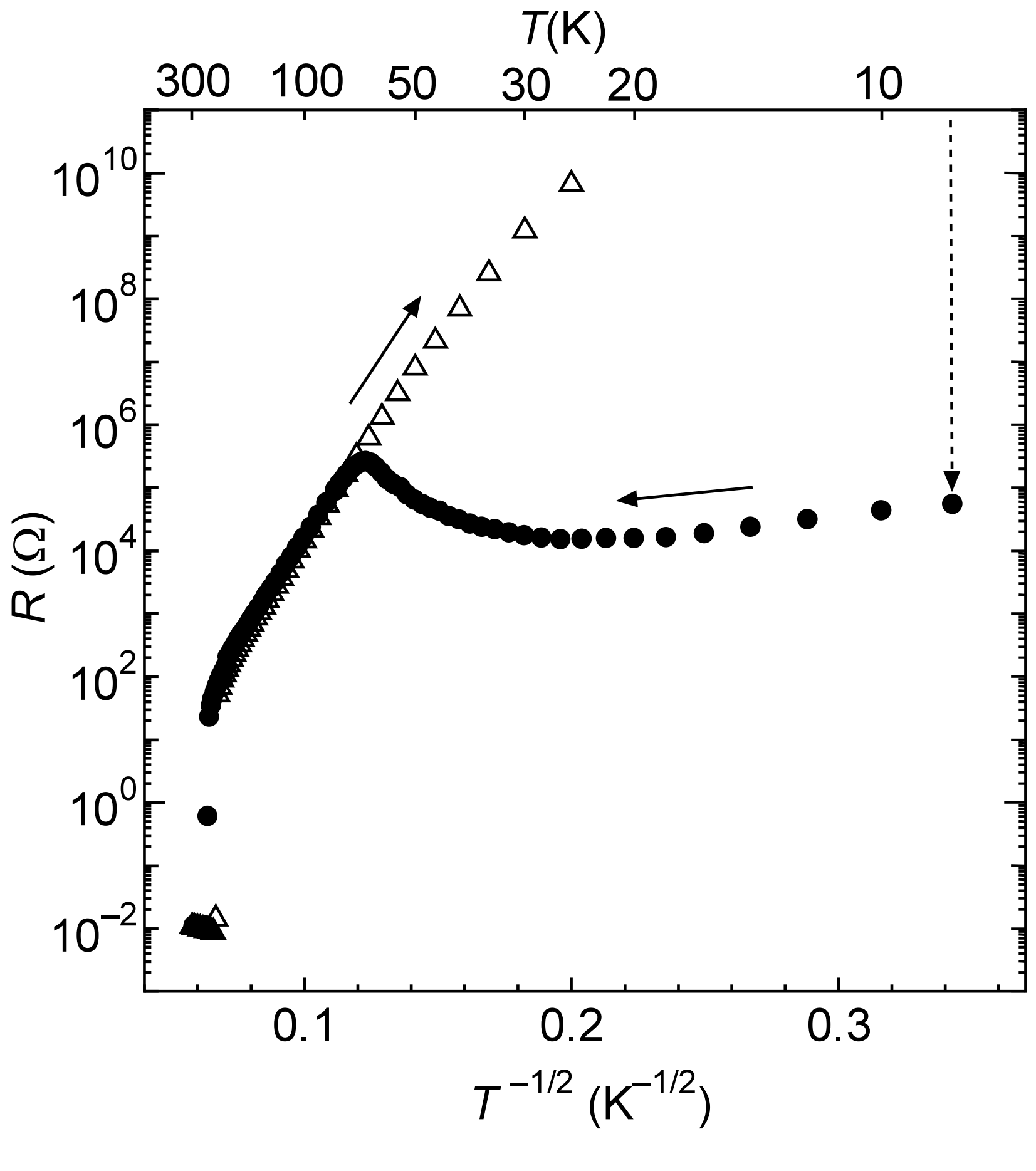

Fig.4, T. Furubayashi

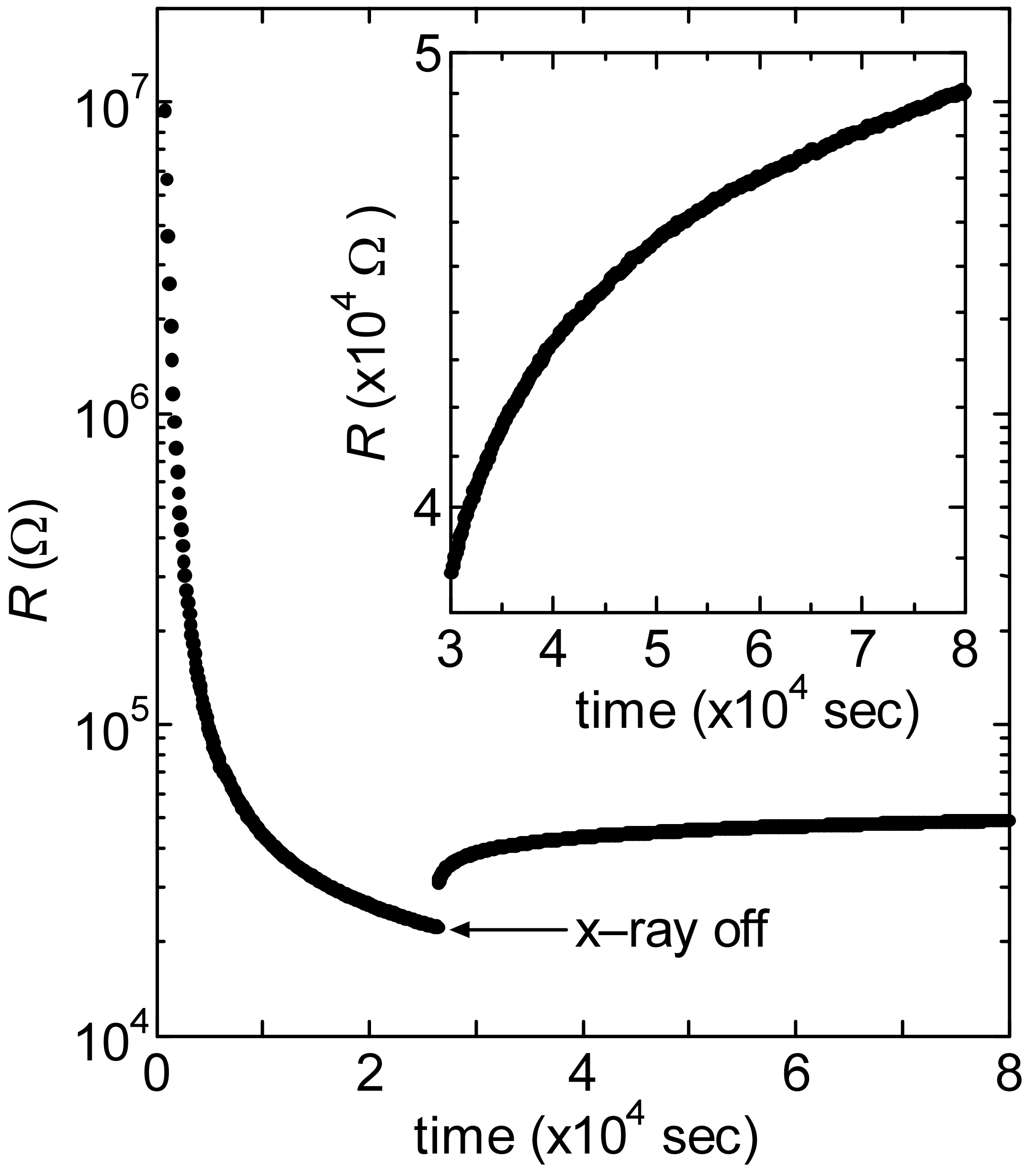

Fig.5, T. Furubayashi

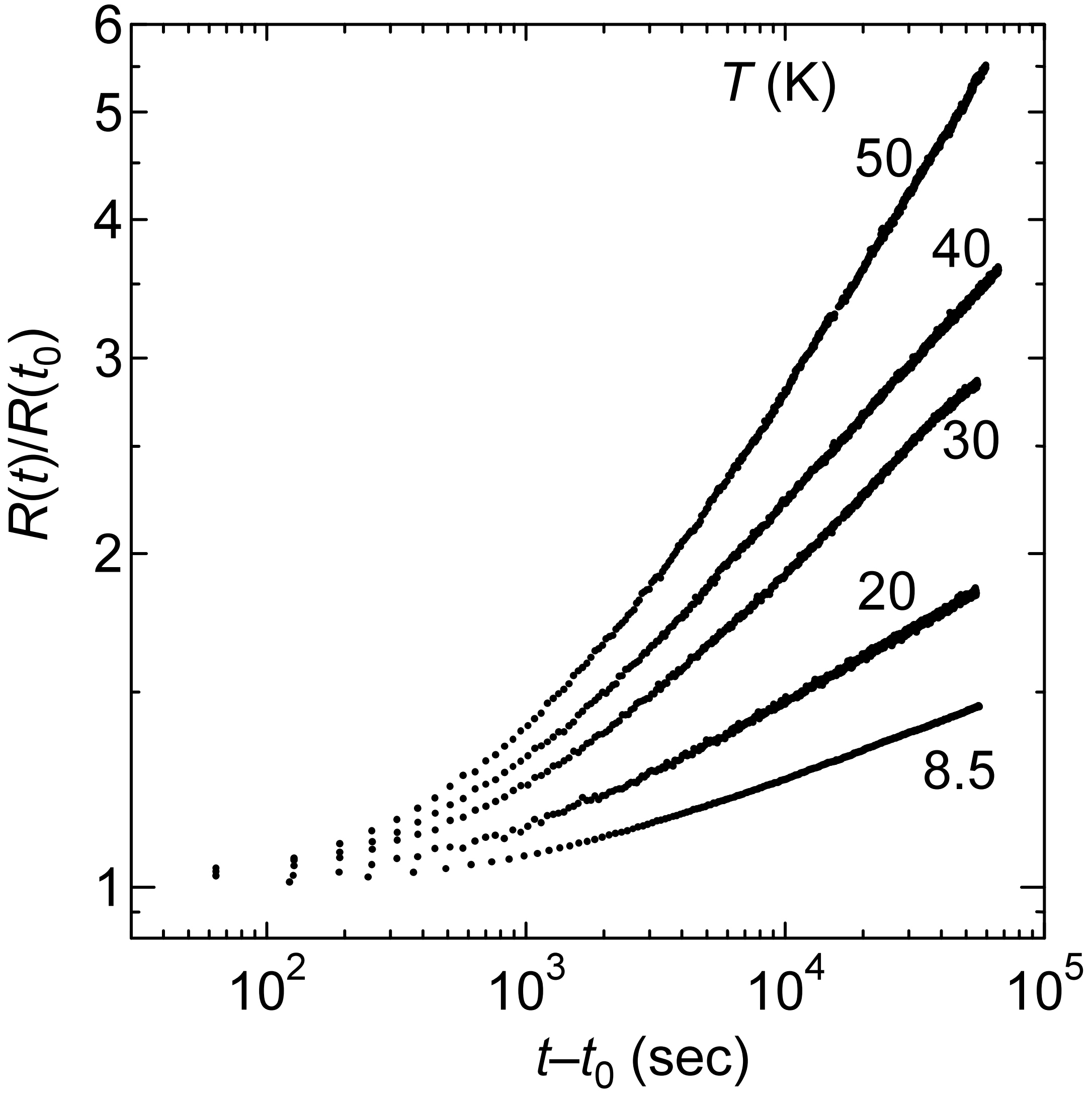

Fig.6, T. Furubayashi